\documentclass[12pt]{article}

\usepackage{latexsym}

\usepackage[T2B]{fontenc}
\usepackage[ctt]{inputenc}

\begin{document}

\title{Singularity free cosmological solutions of Einstein-Maxwell
equations }

\author{
     Stoytcho S. Yazadjiev\thanks{E-mail:
     yazad@phys.uni-sofia.bg}\, ,
     Ventseslav A. Rizov \thanks{E-mail: rizov@phys.uni-sofia.bg}\\
{\footnotesize  Department of Theoretical Physics,
                Faculty of Physics, Sofia University,}\\
{\footnotesize  5 James Bourchier Boulevard, Sofia~1164, Bulgaria }\\
}

\date{}

\maketitle

\begin{abstract}
We report on a new two-parameter class of cosmological solutions
to the Einstein-Maxwell equations. The solutions have everywhere
regular curvature invariants. We prove that the solutions are
geodesically complete and globally hyperbolic.
\end{abstract}


\sloppy
\renewcommand{\baselinestretch}{1.3} %
\newcommand{\sla}[1]{{\hspace{1pt}/\!\!\!\hspace{-.5pt}#1\,\,\,}\!\!}
\newcommand{\db}{\,\,{\bar {}\!\!d}\!\,\hspace{0.5pt}}
\newcommand{\partb}{\,\,{\bar {}\!\!\!\partial}\!\,\hspace{0.5pt}}
\newcommand{\dsla}{\partb}
\newcommand{\eql}{e _{q \leftarrow x}}
\newcommand{\eqr}{e _{q \rightarrow x}}
\newcommand{\ite}{\int^{t}_{t_1}}
\newcommand{\itz}{\int^{t_2}_{t_1}}
\newcommand{\itd}{\int^{t_2}_{t}}
\newcommand{\lfrac}[2]{{#1}/{#2}}
\newcommand{\dV}{d^4V\!\!ol}
\newcommand{\ben}{\begin{eqnarray}}
\newcommand{\een}{\end{eqnarray}}
\newcommand{\la}{\label}

\section{Introduction}

The reasons to study inhomogeneous cosmologies  have deep
observational and theoretical bases. As it is well known the
present universe is not exactly spacially homogeneous even at
large scales. Moreover, there are no reasons to assume that the
regular expansion is suitable for a description of the early
universe. This strongly motivates the study of inhomogeneous
cosmological models. They allows one to investigate a number of
long-standing questions regarding the structure formation, the
occurrence of singularities, the behavior of the solutions in the
vicinity of a singularity and the possibility for our universe to
arise from generic initial data. In particular, on the base  of
some exact solutions it was shown that the nonlinear
inhomogeneities could regularize the initial singularity giving
rise to completely regular cosmologies both in general relativity
\cite{SEN}-\cite{LFJ3} and in alternative gravitational theories
\cite{PIM}-\cite{Y2}. From a purely theoretical point of view, the
investigation of nonsingular cosmological models gives invaluable
insight into the spacetime structure, the inherent nonlinear
character of gravity and its interaction with matter fields. As a
byproduct  it also deepens our understanding of the singularity
theorems, in particular the  assumptions lying in their base
\cite{SEN1}.

In the present work we present a new  two-parameter class of exact
singularity free solutions of the Einstein-Maxwell equations.

\section{The exact solutions}

We consider Einstein-Maxwell gravity described by the equations:

\begin{eqnarray}
R_{\mu\nu}= 2F_{\mu\alpha}F^{\alpha}_{\,\nu} -{1\over 2}g_{\mu\nu}
F_{\alpha\beta}F^{\alpha\beta}, \nonumber \\
\nabla_{\mu}F^{\mu\nu}=0, \\
\nabla_{\alpha}F_{\mu\nu} + \nabla_{\nu}F_{\alpha\mu} +
\nabla_{\mu}F_{\nu\alpha}=0 .\nonumber
\end{eqnarray}

We assume that the spacetime admits two spacelike commuting
Killing vectors $\partial/\partial\phi$ and $\partial/\partial z$,
which are mutually- and hypersurface-orthogonal. The metric then
can be written  in the well known form \cite{KSHMC}:

\begin{equation}
ds^2= e^{2\gamma(t,r)} (-dt^2 + dr^2) +
e^{2\psi(t,r)}\rho^2(t,r)d\phi^2 + e^{-2\psi(t,r)}dz^2.
\end{equation}

We have found the following two-parameter class of exact
solutions:

\begin{eqnarray}
e^{\psi(t,r)} = {\lambda^2 +  (1-\lambda^2)\cosh^4(ar)\cosh^2(2at)
\over \cosh^2(ar)\cosh(2at)}, \nonumber
\end{eqnarray}

\begin{eqnarray}
\rho(t,r)= {1\over a}\cosh(ar)\sinh(ar)\cosh(2at), \nonumber
\end{eqnarray}

\begin{eqnarray}\label{EEMS}
e^{\gamma(t,r)} = \cosh^2(ar)\left[\lambda^2 +
(1-\lambda^2)\cosh^4(ar)\cosh^2(2at) \right],
\end{eqnarray}

\begin{eqnarray}
F_{03} = 4a\lambda \sqrt{1-\lambda^2} e^{-2\psi(t,r)} \tanh(2at),
\nonumber
\end{eqnarray}

\begin{eqnarray}
F_{13} = 4a\lambda \sqrt{1-\lambda^2} e^{-2\psi(t,r)} \tanh(ar),
\nonumber
\end{eqnarray}

where $a>0$ and $0<\lambda<1 $ are free parameters. The limiting
cases $\lambda=0$ and $\lambda=1$ correspond to solutions of the
vacuum Einstein equations. The limit $a\rightarrow 0$ gives the
Minkowski spacetime. The range of the coordinates is:

\begin{equation}
-\infty<t,z<\infty \,\,\, , 0<r<\infty, \,\,\, 0\le \phi  <2\pi.
\end{equation}

The spacetimes described by the solutions (\ref{EEMS}) have well
defined axis of symmetry \cite{KSHMC}. Therefore, it can be said
that these spacetimes admit cylindrical symmetry.

It should be noted that new solutions with the same spacetime
geometry and different Maxwell tensor can be obtained by a duality
rotation

\begin{equation}
F_{\mu\nu} \rightarrow F_{\mu\nu}\cos(\theta) + \star F_{\mu\nu}
\sin(\theta),
\end{equation}

where $\theta$ is a constant parameter and $\star$ is the Hodge
dual.

\section{Curvature invariants}

The components of the Weyl tensor in the standard null tetrad are:

\begin{eqnarray}
e^{2\gamma(t,r)}\Psi_{0}= {3a^2\over \cosh^2(ar)} -2a^2 \left[1-
{\lambda^2\over S(t,r)} \right] \left[{2\over \cosh^2(2at)} +
{1\over \cosh^2(ar)} \right]\nonumber  \\ - {8a^2\lambda^2\over
S(t,r)} \left[1 - {\lambda^2\over S(t,r)} \right]
\left[\tanh^2(2at)+ \tanh^2(ar)\right] \\ +16a^2 \left[1-
{\lambda^2\over S(t,r)}
\right]^2\left[\tanh(2at) + \tanh(ar) \right]^2 \nonumber  \\
- 8a^2 \left[1- {\lambda^2\over S(t,r)} \right] \left[ \tanh(2at)+
\tanh(ar)\right]\tanh(2at) \nonumber
\end{eqnarray}

\begin{eqnarray}
\Psi_{1}=0,
\end{eqnarray}

\begin{eqnarray}
3e^{2\gamma(t,r)}\Psi_{2}= {3a^2 \over \cosh^2(ar)} + 2a^2\left[1-
{\lambda^2\over S(t,r)} \right] \left[{1\over \cosh^2(ar)} -
{2\over \cosh^2(2at)} \right]\nonumber  \\ + {8a^2\lambda^2\over
S(t,r)} \left[1- {\lambda^2\over S(t,r)} \right]
\left[\tanh^2(ar)- \tanh^2(2at)\right] \\
+ 16a^2 \left[1- {\lambda^2\over S(t,r)} \right]^2
\left[\tanh^2(2at) - \tanh^2(ar)\right]    \\
+4a^2 \left[1- {\lambda^2\over S(t,r)} \right]\left[3\tanh^2(ar) -
2\tanh^2(2at) -1 \right]
\end{eqnarray}

\begin{eqnarray}
\Psi_{3}=0,
\end{eqnarray}

\begin{eqnarray}
e^{2\gamma(t,r)}\Psi_{4}= {3a^2\over \cosh^2(ar)} -2a^2 \left[1-
{\lambda^2\over S(t,r)} \right] \left[{2\over \cosh^2(2at)} +
{1\over \cosh^2(ar)} \right]\nonumber  \\ - {8a^2\lambda^2\over
S(t,r)} \left[1 - {\lambda^2\over S(t,r)} \right]
\left[\tanh^2(2at)+ \tanh^2(ar)\right] \\ +16a^2 \left[1-
{\lambda^2\over S(t,r)}
\right]^2\left[\tanh(2at) - \tanh(ar) \right]^2 \nonumber  \\
- 8a^2 \left[1- {\lambda^2\over S(t,r)} \right] \left[ \tanh(2at)
- \tanh(ar)\right]\tanh(2at) \nonumber
\end{eqnarray}

where $S(t,r)=\lambda^2+(1-\lambda^2)\cosh^4(ar)\sinh^2(2at)$.

Further, the components of the Maxwell tensor in the standard null
tetrad read

\begin{eqnarray}
\Phi_{0}={-i\over S(t,r)}2a\lambda \sqrt{ 1-\lambda^2}
\cosh(2at)[\tanh(2at)-\tanh(ar)],
\end{eqnarray}

\begin{eqnarray}
\Phi_{1}=0,
\end{eqnarray}

\begin{eqnarray}
\Phi_{2}={i\over S(t,r)}2a\lambda \sqrt{1-\lambda^2})
\cosh(2at)[\tanh(2at)+\tanh(ar)].
\end{eqnarray}

All components of the Weyl tensor and the Maxwell tensor are
regular everywhere. Since the scalar invariants that can be formed
with the metric and the Riemann curvature tensor are polynomials
of these components,  all curvature invariants are regular
everywhere. From the explicit form of the Weyl tensor components
one can check  that the spacetimes are of Petrov type $I$ except
at the axis where they are of type $D$.

\section{Geodesic completeness}

In order to demonstrate the geodesic completeness of the above
solutions we have to show that all causal geodesics can be
extended to arbitrary values of the affine parameter. Since the
metric functions  are even in the time variable we shall
investigate only future-directed geodesics. Below we consider
$0<\lambda<1$.

The existence of isometries gives rise to two constants of motion
along the geodesics:

\begin{eqnarray}
L= e^{2\psi(t,r)}\rho^2(t,r){d\phi\over d\tau}, \nonumber\\
P= e^{-2\psi(t,r)}{dz\over d\tau},
\end{eqnarray}
where we have denoted by $\tau$ the affine parameter along the
geodesics.

The affinely parameterized geodesics satisfy

\begin{equation}\label{APG}
e^{2\gamma(t,r)}\left[\left({dt\over d\tau}\right)^2 -
\left({dr\over d\tau}\right)^2 \right] - {L^2\over
\rho^2(t,r)}e^{-2\psi(t,r)} - P^2 e^{2\psi(t,r)} = \epsilon,
\end{equation}

where $\epsilon=1$ and $\epsilon=0$ for timelike and null
geodesics, respectively. Writing $d\phi/ d\tau$ and $dz/ d\tau$ as
functions of $L$ and $P$, the geodesic equations for $t$ and $r$
can be written in the following form \cite{LFJ2},\cite{LFJ3}:

\begin{eqnarray}
{d \over d\tau}\left(e^{2\gamma(t,r)}{dt\over d\tau} \right)=
e^{-2\gamma(t,r)}M(t,r)\partial_{t}M(t,r), \\
{d \over d\tau}\left(e^{2\gamma(t,r)}{dr\over d\tau} \right)= -
e^{-2\gamma(t,r)}M(t,r)\partial_{r}M(t,r),
\end{eqnarray}

where the function $M(t,r)$ is defined by

\begin{equation}
M(t,r) = e^{\gamma(r,r)}\left[\epsilon + {L^2 \over
\rho^2(t,r)}e^{-2\psi(t,r)} + P^2 e^{2\psi(t,r)} \right]^{1/2} .
\end{equation}

First we consider the null geodesics with $L=P=0$. For this case
we have $dt/d\tau=|dr/d\tau|$ and

\begin{equation}
{d\over d\tau}\left(e^{2\gamma(t,r)}{dt\over d\tau}  \right)=0.
\end{equation}

After integration, we obtain

\begin{equation}
{dt\over d\tau}= C e^{-2\gamma(t,r)},
\end{equation}

where $C>0$ is a constant. One can easily see from the explicit
form of $e^{\gamma(t,r)}$ that $e^{-2\gamma(t,r)}\le 1 $. As a
consequence one finds

\begin{equation}
{dt\over d\tau}= |{dr\over d\tau}| \le C
\end{equation}

and therefore the null geodesics with $L=P=0$ are complete.

Let us turn to the general case when at least one of the constants
$\epsilon$, $L$ or $P$ is different from zero. In this case we
parameterize $dt/d\tau$ and $dr/d\tau$ by writing
\cite{LFJ2},\cite{LFJ3}:

\begin{eqnarray}
{dt\over d\tau}= e^{-2\gamma(t,r)}M(t,r)\cosh(\upsilon),\nonumber \\
{dr\over d\tau}= e^{-2\gamma(t,r)}M(t,r)\sinh(\upsilon).
\end{eqnarray}

Substituting these expressions into the equations for $dt/d\tau$
and $dr/d\tau$, we obtain the following equation for $\upsilon$:

\begin{equation}
{d\upsilon\over d\tau}= - e^{-2\gamma(t,r)}
\left[\partial_{t}M(t,r)\sinh(\upsilon) +
\partial_{r}M(t,r)\cosh(\upsilon) \right].
\end{equation}

The above equation can be written in a more explicit form

\begin{equation} \label{DV1}
{d\upsilon\over d\tau}= -{1\over M(t,r)} \{
J_{+}(t,r)\cosh(\upsilon) + J_{-}(t,r)\sinh(\upsilon) \},
\end{equation}

where

\begin{eqnarray}\label{DV2}
J_{+}(t,r) = 2a\epsilon \tanh(ar)\left[1 +
{2(1-\lambda^2)\cosh^4(ar)\cosh^2(2at) \over \lambda^2 +
(1-\lambda^2)\cosh^4(ar)\cosh^2(2at)  } \right] \nonumber \\
 + {a^3L^2[3\tanh^2(ar)-1] \over \tanh^3(ar)[\lambda^2 +
(1-\lambda^2)\cosh^4(ar)\cosh^2(2at) ]} \\
+ 8aP^2 (1-\lambda^2)[\lambda^2 +
(1-\lambda^2)\cosh^4(ar)\cosh^2(2at)]\tanh(ar)\nonumber ,
\end{eqnarray}

\begin{eqnarray}\label{DV3}
J_{-}(t,r) = 4a\epsilon{(1-\lambda^2)\cosh^4(ar)\cosh^2(2at)
\over\lambda^2 + (1-\lambda^2)\cosh^4(ar)\cosh^2(2at)  }  \nonumber \\
+ 2aP^2 {[\lambda^2 + (1-\lambda^2)\cosh^4(ar)\cosh^2(2at) ] \over
\cosh^4(ar)\cosh^2(2at)}  \\
\times \left[3(1-\lambda^2)\cosh^4(ar)\cosh^2(2at) - \lambda^2
\right]\tanh(2at) \nonumber .
\end{eqnarray}

In order for the geodesics to be complete, the functions
$dt/d\tau$ and $dr/d\tau$ have to remain finite for finite values
of the affine parameter. In fact, it is sufficient to consider
only $dt/d\tau$ since $dt/d\tau$ and $dr/d\tau$ are related via
(\ref{APG}) and $dr/d\tau$ cannot become singular if $dt/d\tau$ is
not singular. The derivatives $d\phi/d\tau$ and $dz/d\tau$ are
regular functions of $t$ and $r$ and cannot become singular if
$t(\tau)$ and $r(\tau)$ are not singular. The only problem we
could have is when $r(\tau)$ approaches the value $r=0$ for $L\ne
0$. We shall show, however, that $r(\tau)$ cannot become zero for
$L\ne 0$.

First we consider the geodesics with increasing $r$ (i.e.,
$\upsilon >0$). From the explicit form of the functions
$e^{2\gamma(t,r)}$ and $M(t,r)$ it is not difficult to see that

\begin{equation}
e^{-2\gamma(t,r)}M(t,r) \le \left[ \epsilon + P^2 + {a^2L^2\over
\sinh^2(ar)} \right]^{1/2}.
\end{equation}

Therefore, $dt/d\tau$ could become singular only because of
$\upsilon(\tau)$. However, for increasing $r$, $\upsilon(\tau)$
cannot diverge since for large $t$ (large $r$) the derivative
$dt/d\tau$ becomes negative as can be seen from (\ref{DV1}),
(\ref{DV2}) and (\ref{DV3}).

Let us now consider the second case when $r(\tau)$ decreases
($\upsilon$ <0). The geodesics with $L=0$ reach the axis $r=0$
smoothly and then reappear with $dr/d\tau>0$ ($\upsilon>0$). A
problem may arise from $r=0$ for $L\ne 0$. However, for $L\ne 0$,
$r(\tau)$ cannot become zero and this can be shown as follows.
When $r(\tau)$ approaches zero the dominant term is the one
associated with $L$ and the other terms can be ignored. So, for
very small $r$ the geodesics behave as null geodesics with $P=0$:

\begin{eqnarray}
{dt\over d\tau}= e^{-2\gamma(t,r)}M(r)\cosh(\upsilon) ,\nonumber \\
{dr\over d\tau}= e^{-2\gamma(t,r)}M(r)\sinh(\upsilon) ,\\
{d\upsilon\over d\tau}= - e^{-2\gamma(t,r)}
\partial_{r}M(r)\cosh(\upsilon) ,\nonumber
\end{eqnarray}

where

\begin{equation}
M(r)= |aL| {\cosh^3(ar)\over \sinh(ar)}.
\end{equation}

Hence, one finds

\begin{equation}
{dr\over d\upsilon}= - {M(r)\over
\partial_{r}M(r)}\tanh(\upsilon).
\end{equation}

After integration we have

\begin{equation}
\sinh(ar)=D \cosh^3(ar)\cosh(\upsilon),
\end{equation}

where $D>0$ is a constant. From here one can immediately see that
$r(\tau)$ cannot become zero. So,  we have proven that our
solutions are geodesically complete. In the same manner it can be
shown that the vacuum solutions for $\lambda=0$ and $\lambda=1$
are geodesically complete, too.

From the above considerations it follows that every maximally
extended null geodesic intersects once and only once any of the
hypersurfaces $t=constant$. Therefore the hypersurfaces
$t=constant$ are global Cauchy surfaces \cite{Geroch} and the
spacetimes described by the solutions are globally hyperbolic.

Finally, it is interesting to see which of the assumptions of the
singularity theorems \cite{HE} turn out to be violated. Since the
energy  and the causal conditions are fulfilled it remains to
conclude that the spacetimes considered do not contain a closed
trapped surface. In order to prove this we shall follow  the
considerations of \cite{CFJS},\cite{SEN2} and \cite{SEN3}. Indeed,
let us suppose that the there is one such surface. Then, since the
surface is compact, there must be a point  $q$ where $r$ reaches
its maximum. Let us denote it by $r_{max}=R$ on a constant time
hypersurface $t=T$. For the traces of both null second fundamental
forms at $q$, it can be shown that

\begin{eqnarray}
K^{+} \ge \sqrt{2}a e^{-\gamma(R,T)} {1+ \tanh(2at)\tanh(2ar)
\over \tanh(2ar)} >0 ,\nonumber \\
K^{-} \le \sqrt{2}a e^{-\gamma(R,T)} {\tanh(2at)\tanh(2ar)-1 \over
\tanh(2ar)} <0.
\end{eqnarray}

The traces have opposite  sings, and, therefore, there are no
closed trapped surfaces.

In conclusion, we have presented a new (to the best of our
knowledge) two-parameter class of exact solutions of the
Einstein-Maxwell equations. The solutions have no curvature
singularity. Moreover, they are geodesically complete and globally
hyperbolic. The solutions  can be viewed as explicit examples of
how the nonlinear inhomogeneities can regularize the singularities
yielding completely regular spacetimes.

\end{document}